\documentstyle[epsfig]{mn2e}
\begin{document}

\title
[The Galaxy luminosity Function]  
{The Galaxy Luminosity Function from $M_R = -25$ to $M_R = -9$.}

\author[Trentham et al.]  
{
Neil Trentham$^{1}$, Leda Sampson$^{1}$ \& Manda Banerji$^{1}$\\ 
$^1$ Institute of Astronomy, Madingley Road, Cambridge, CB3 0HA.\\ 
}
\maketitle 

\begin{abstract} 
        {Redshift surveys like the Sloan Digital Sky Survey (SDSS) have given
        a very precise measurement of the galaxy luminosity function
        down to about $M_R = -17$ ($\approx M_B = -16$).
        Fainter absolute magnitudes cannot be probed because of the flux limit
        required for spectroscopy. Wide-field surveys of nearby
        groups using mosaic CCDs on large telescopes are able to reach
        much fainter absolute magnitudes, about $M_R = -10$. These
        diffuse, spiral-rich groups are thought to be typical 
	environments for galaxies so
        their luminosity functions should be
        the same as the field luminosity function. The luminosity function
	of the groups at the bright end ($M_R < -17$) is limited by Poisson
	statistics and is far less precise than that 
	derived from redshift surveys.
	Here we combine the
        results of the SDSS and the surveys of nearby groups
        and supplement the results with studies of Local Group
        galaxies in order to determine the galaxy luminosity function
        over the entire range $-25 <M_R < -9$.  The average
        logarithmic slope of the field luminosity function 
	between $M_R = -19$ and $M_R = -9$ is $\alpha = -1.26$, although
	a single power law is a poor fit to the data over the
	entire magnitude range. 
        We also determine the luminosity function of galaxy
        clusters and demonstrate that it is different from the field
        luminosity function at a high level of significance: there are
        many more dwarf galaxies in clusters than in the field, due to
        a rise in the cluster luminosity function of $\alpha \sim
        -1.6$ between $M_R = -17$ and $M_R = -14$.

}
\end{abstract} 


\section{Introduction} 

        The galaxy luminosty function (LF) $\phi(L)$ is defined such that
        $\phi(L) \, {\rm d}L$ is the density of galaxies having
        luminosities between L and L + ${\rm d}L$. It is usually
        regarded as a convenient test of galaxy formation theories
        because it is so easy to measure.  However, the observed galaxy LF
 	is different from the galaxy mass function,
        which is more directly predicted by theory.
        This difference is particularly significant in the
        context of cold dark matter galaxy formation theories
        (Moore et al.~1999;
        Klypin et al.~1999; Bullock,
        Kravstov \& Weinberg  2001;
        Chiu, Gnedin \& Ostriker 2001; Benson et al.~2003a,b;
        Kazantzidis et al.~2004).
	
        There are a number of redshift surveys that have
        recently been completed. These surveys measure redshifts and
        therefore distances for large numbers of galaxies in 
        flux-limited samples. 
        The largest such survey is the Sloan Digital Sky Survey 
	(SDSS; www.sdss.org)
        which has measured redshifts for $>10^5$ galaxies. The LF
        produced by this survey has very small uncertainties. The
        Poisson errors are negligible due to the sample size, and the
        sample is complete because the vast majority of high-luminosity
	galaxies have
        high surface brightnesses (Cross et al.~2001)
        and so are contained within the sample. 
        A number of other surveys (see Section 2) also have produced
        accurate LFs, all of which are consistent with each other
        given the uncertainties that 
	come from matching the different filters used
        in the different surveys and from cosmic variance. 
        The SDSS redshift sample is only complete to absolute magnitudes of
        about $M_R = -17.5$. This limit is imposed by the requirement
        that galaxies need to be bright enough for spectroscopy. The
        spectroscopy limit is quite bright because the
        SDSS uses a relatively small (2.5 m) telescope. 

        Another recent development has been the advent of mosaic CCDs
        that can be used 
        on large telescopes (e.g.~SuprimeCam on the NAOJ Subaru 8 m
        Telescope; Miyazaki et al.~2002). 
	This has permitted observations of large
        areas of sky down to very faint limits.  
        A number of low-surface-brightness galaxies are seen in
	fields within galaxy groups but not in blank fields.
        These are interpreted as dwarf galaxies
        in the group, not background galaxies (Flint et al.~2001a).
        We can therefore infer their
        distances without spectroscopy, which is not possible for these faint
        low-surface-brightness galaxies.
        Hence measurements of the LF in nearby ($<$ 20 Mpc)
        groups are possible down to $M_R = -10$. 

        Diffuse, spiral-rich groups are typical
        environments for most galaxies.
        Therefore it should be
        possible to append the LFs of these groups to the field LF from
        SDSS and obtain the galaxy LF over a very wide magnitude
        range. This is the subject of the present paper.

        At the very faint end ($M_R \sim -10$), the surveys of groups
        run into difficulty because of the inability to assign
        memberships to faint low-surface-brightness galaxies.  At $R
        > 22$, low-surface-brightness ($\mu >25 \, R$ mag arcsec$^{-2}$
        over 2 or more seeing disks) 
        galaxies appear in
        blank field surveys.  Thus, while excess numbers of 
        low-surface-brightness galaxies may be
        seen in the group fields, we can only infer group 
        membership in a statistical
        sense.  For example, none of the candidate galaxies with $M_R > -11$  
        in the diffuse groups studied by Trentham \& Tully (2002) can
        be regarded as highly probable members on an individual basis. 

        The only environment where the LF can be measured 
        fainter than $M_R = -10$ is the Local Group (the faintest galaxy
        in the Local Group, Andromeda IX, has $M_R = -9$; Zucker et al.~2004). 
	We therefore 
        supplement the combined SDSS+Groups
	LF with the Local Group LF at the
        very faint end. 
        
        The paper is structured as follows. In Section 2 we describe
        results from redshift surveys. In
        Section 3 we investigate the behaviour of the LF at very
        bright magnitudes ($M_R < -23$).
        In Section 4 we describe
        results from deep wide-field surveys of groups. In Section 5
        we compute the $R$-band Local Group LF.
        In Section 6 we combine all the
        individual LFs to determine the galaxy LF between $M_R = -25$
        and $M_R = -9$. In Section 7 we investigate analytic forms
        that describe this function.
        In Section 8 we do a brief comparison with
        theory.  In Section 9 we compute the
        cluster LF and investigate how it is different from the field
        LF. Finally in Section 10 we summarize. Throughout this work,
        we assume the following cosmological parameters: $H_0 = 65\,\,
        {\rm km} {\rm s}^{-1} {\rm Mpc}^{-1}$, $\Omega_{\rm
        matter}=0.3$, $\Omega_{\Lambda}=0.7$.

\section{Results from redshift surveys}

The SDSS LF is most appropriate for this study because 
its $r^{\prime}$ filter is similar 
to the $R$ filter that was used in CCD mosaic studies of nearby
groups.  
It is derived from measurements of $>10^{5}$ galaxies so 
the statistical uncertainties are small. There are a 
number of other redshift surveys which have
similarly produced field LFs (Table 1).

        The SDSS LF is described in detail by Blanton et al.~(2001,
        2003). In Figure 1 we present the SDSS LF
        converted to the $R$ magnitude system that we use in
        this paper.

\begin{table*}
\caption{Spectroscopic field surveys}
{\vskip 0.75mm}
{$$\vbox{
\halign {#\hfil && \quad #\hfil \cr
\noalign{\hrule \medskip}
Survey & Reference & Filter & Limiting Magnitude\cr
\noalign{\smallskip \hrule \smallskip}
\cr
Stromolo-APM    & Loveday et al.~1992  &  $b_J$ &       $-17$ &\cr
Hawaii-Caltech  & Cowie et al.~1996    &  $B$   &       $-12.5$ &\cr
                &                      &  $K$    &       $-20.5$ &\cr
Autofib         & Ellis et al.~1996    &  $b_J$ &       $-15$ &\cr
LCRS            & Lin et al.~1996      &  $R$   &       $-18$ &\cr
ESO Slice Project& Zucca et al.~1997  &  $b_J$ &       $-15$ &\cr
SDSS            & Blanton et al.~2001, 2003   &  $u^{\prime}$   &       $-16.5$ &\cr
                &                             &  $g^{\prime}$   &       $-17.5$ &\cr
                &                             &  $r^{\prime}$   &       $-17$ &\cr
                &                             &  $i^{\prime}$   &       $-18$ &\cr
                &                             &  $z^{\prime}$   &       $-19.5$ &\cr
2MASS +2df &  Cole et al.~2001 &  $J$         &  $-19$ &\cr
                &              &  $K_s$       &  $-20$ &\cr
2df &  Norberg et al.~2002 &  $b_J$    &       $-17.5$ &\cr
\noalign{\smallskip \hrule}
\noalign{\smallskip}\cr}}$$}
\end{table*}

\begin{figure}
\begin{center}
\vskip-4mm
\psfig{file=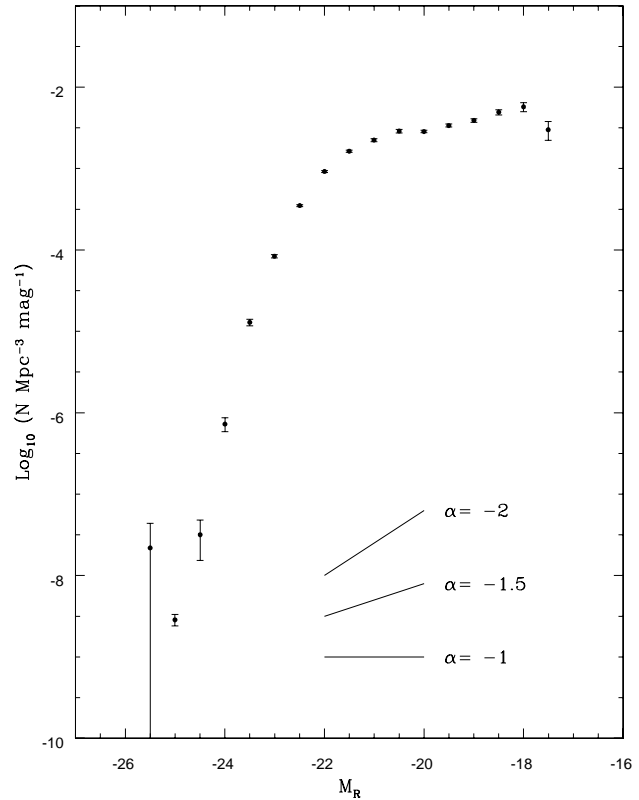, width=8.65cm}
\end{center}
\vskip-3mm
\caption{
The SDSS luminosity function, from Blanton et al.~(2003), corrected to
our cosmology.
This is derived
from the SDSS $r^{\prime}$ LF assuming a colour correction of $r^{\prime} - R$
varying from 0.24 at $M_R = -24$ to 0.17 at $M_R = -16$.
These conversions come from
the caliibrations of Fukugita, Shimasaku and Ichikawa (1995)
and the colour-dependent LFs of Blanton et al.~(2001). Three
values of $\alpha = {\rm d}\,{\rm log}\phi(L)/ {\rm d}\,{\rm log} L$,
the logarithmic slope of the luminosity function, are shown.  }
\end{figure}

\section{The very bright end}

At the very bright end the SDSS LF 
is not a direct measurement. This is because most of the
galaxies in the SDSS sample with $M_R < -23$ are
sufficiently distant that evolutionary corrections 
to redshift zero are significant.
Figure 2 shows that at absolute magnitudes
$M_R < -23$ the majority of galaxies have redshifts $z > 0.2$,
where corrections are typically 0.3 mag
(Lin et al.~1999; Blanton et al.~2003).

\begin{figure}
\begin{center}
\vskip-4mm
\psfig{file=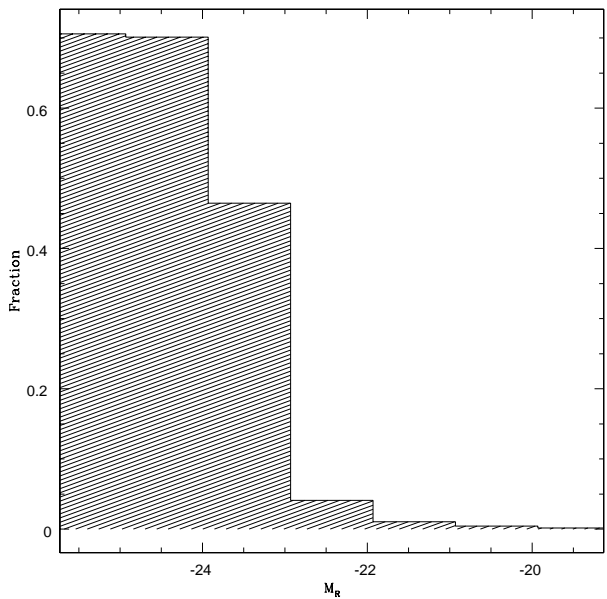, width=8.65cm}
\end{center}
\vskip-3mm
\caption{
The fraction of galaxies at redshift $z > 0.2$ in the SDSS sample used
to compile Figure 1.
}
\end{figure}
        
There are too few galaxies in the local Universe ($D < 40$ Mpc;
{\it Nearby Galaxies Catalog}, Tully 1988)  
for this region of the LF to be measured directly: only one galaxy (the Seyfert
galaxy NGC 4594) has $M_R < -24$ and 25 have $M_R < -23$.
For $D < 40$ Mpc, the ratio of galaxies with $-24 < M_R < -23.5$ to those
with $-23.5 < M_R < -23$ is
$0.20 \pm 0.11$.  
The LF shown in Figure 1 implies a value for this ratio of 0.09,
consistent with the number for the local Universe number.
This concordance suggests that
the evolutionary corrections made to the
SDSS data, which are computed using observations of galaxies
whose absolute magnitudes span a wide range, are appropriate
at the very bright end $M_R < -23$.

\section{Results from deep Mosaic CCD surveys}

In Table 2 we list some nearby groups where deep CCD mosaic
imaging has been performed. 
The LFs from these surveys extend down to $M_R = -10$. Earlier
photographic studies (e.~g.~Ferguson \& Sandage
1991) covered larger areas but did not reach magnitudes
this faint. 
All of these groups have been imaged in
the Cousins $R$ band ($\lambda_{\rm eff} = 6588\, \AA$),
which is close to the SDSS $r^{\prime}$ band ($\lambda_{\rm eff} =
6290\, \AA$; Fukugita et al.~1995).
The $R$-band LF of these groups is therefore
suitable for extending the SDSS LF to faint magnitudes.
        
\begin{table*} \caption{Groups with deep luminosity function measurements}
{$$\vbox{
\halign {#\hfil && \quad #\hfil \cr
\noalign{\hrule \medskip}
Group & Distance & limiting $M_R$ & reference & survey telescope & sample size  &\cr
 & Mpc & & & & &\cr
\noalign{\smallskip \hrule \smallskip}
\cr
Ursa Major      & 18.6 & $-11$ & Trentham et al.~2001a (TTV01) & CFHT  & 50  &\cr
Leo I           & 11.1 & $-10$ & Flint et al.~2001b & KPNO 0.9 m & 112 &\cr
Coma I          & 16.4 & $-10$ & Trentham \& Tully 2002 (TT02) & Subaru  & 38 &\cr
NGC 1023 Group  & 10.0 & $-10$ & Trentham \& Tully 2002 (TT02) & Subaru & 28  &\cr
\noalign{\smallskip \hrule}
\noalign{\smallskip}\cr}}$$}
\end{table*}

Members are identified based on their surface
brightnesses. Low-surface-brightness galaxies that appear within
a group of giant galaxies and whose counterparts do not
exist in blank fields are assumed to lie at the distance of
the group.
Distances are therefore determined based on
morphology; spectroscopic redshifts are
unattainable for these faint low-surface-brightness galaxies.
At the faintest magnitudes this exercise
becomes statistical as a few very small low-surface-brightness
galaxies are visible in blank fields. We tend to see more such
galaxies in fields containing nearby groups of large galaxies,
but for any particular galaxy we can only assign membership on
probabilistic grounds.

\begin{figure}
\begin{center}
\vskip-4mm
\psfig{file=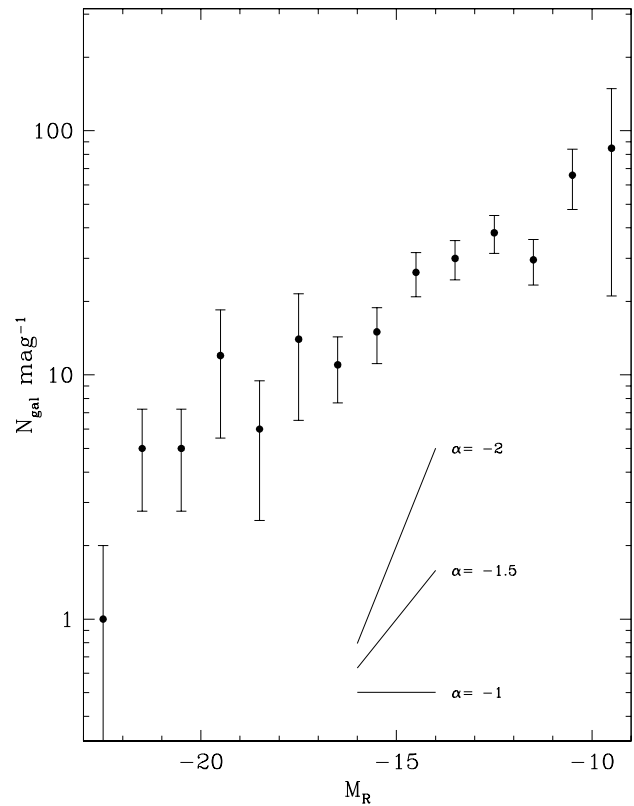, width=8.65cm}
\end{center}
\vskip-3mm
\caption{
The composite luminosity function for the groups
listed in Table 2, with errors computed as decribed in
the text.  Here $N_{\rm gal}$ is the total number of galaxies in
the combined sample, proportional to $N$ in Figure 1.
}
\end{figure}

The composite LF of the groups in Table 2 is presented in
Figure 3.  Computing this was not straightforward because of the
different methods used by different authors to assess membership
probabilities.  For Ursa Major, Coma I, and the  
NGC 1023 Group (hereafter G1), TTV01 and TT02 employed a subjective rating
system based on surface brightnesses, morphologies and
spectroscopy (when available).
For the present calculation, galaxies were 
assigned a high (rating ``0--2'' in the notation of TTV01 and TT02)
or low (rating ``3'') probability of membership.
For the Leo I Group (hereafter G2), Flint, Bolte \& Mendes de Olivera (2001b) computed
completeness corrections $C(M)$in each magnitude interval.
The number of galaxies in the
composite group sample in each magnitude interval equals
the number of G1 galaxies with a high membership probability
plus 1/2 times the
the number of 
G1 galaxies with a low membership probability plus $(1+C(M))$ times
the number of galaxies in the
G2 sample. 
The weighting of 1/2 for the G1 galaxies with a low membership probability
comes from a study of the radial distribution of galaxies with this 
rating within the groups (TTV01, TT02).

The uncertainties in these numbers are potentially large because the
rating system used by TTV01 and TT02 is subjective and 
the completeness corrections used for the G2 sample are model-dependant. 
Errors in the number of group members will be systematic and a
conservatve approach will be required to
determine the uncertainties; we
estimate them from the quadrature sum of Poisson errors
(Gehrels 1986), the entire contribution of G1 galaxies with 
a low membership probability  
and the entire enhancement in the number of G2 galaxies due to 
completeness corrections.  That the error bars in Figure 3 are so large
at the faintest magnitudes follows from the fact that 
the contributions from the last two terms dominate the uncertainties. 

        The LF of the groups over the range $M_R=-19$ to $M_R=-10$ is
        fit by a power law of index $\alpha = -1.26 \pm
        0.11$.  This is an interesting magnitude range because
        $M_R = -19$ ($M_B \sim -18$) is the brightest magnitude where
	dwarf galaxies, as identified by their position on a
        magnitude--surface-brightness plot (e.~g.~Figure 1 of Binggeli
	1994), contribute significantly to the LF.  

Precise distances have been measured to individual 
low-luminosity galaxies in Centaurus and Sculptor (Jerjen,
Freeman \& Binggeli 2000; Karachentsev et al.~2002). 
These groups will be good 
candidates to add to the list in Table 2
when complete galaxy samples are available; this
will reduce the errors in Figure 3.

\section{The Local Group}

The LF of groups
has large errors at $M_R \sim -10$ 
which can be decreased
by including
the Local Group in the analysis. 
In the Local Group, the lowest-luminosity galaxies can 
be resolved into stars so that membership
can be established with confidence.                
However, there are very few galaxies 
and the Poisson errors  
are large (but not significantly larger than the errors in the
LF of groups at $M_R \sim -10$).

The faintest galaxies
in the Local Group would have been detected in the nearby groups studied 
by Trentham \& Tully 2002 given the detection limit of that survey
(1$\sigma$ of about 29 $R$ mag arcsec$^{-2}$).  However
only a small part of the galaxies would be 
visible above the sky and the galaxies would be
indistinguishable from a large number of small background
galaxies, many of which have been dimmed by $(1+z)^4$ cosmological effects.
The ability to distinguish members from background galaxies can only be
performed with any confidence brighter than $M_R = -10$.

There are, however, complications in deriving a Local Group LF that can
used in the current analysis. 

        Firstly there are uncertainties as to the membership of stars
        in specific galaxies. For example, in Draco, inclusion of a
        number of new stars caused the tidal radius to increase
        significantly (Piatek et al.~2001
	Odenkirchen et al.~2001) which in turn caused the absolute
        magnitude to brighten. Therefore absolute 
	magnitudes of individual Local 
        Group galaxies are uncertain.

        Secondly, most Local Group galaxies have not been observed in
        the $R$-band.  Therefore we need to perform a colour correction
	for each dwarf.

Thirdly, the Local Group sample may be incomplete at the faint
end.  The lowest-luminosity galaxies have very low surface
brightnesses and so can be difficult to find.
They are particularly difficult to find if they are viewed in
projection against either the Milky Way or
the M31 disks (e.~g.~Andromeda IX; Zucker et al.~2004).
They are also difficult to find if they are in the process of
being disrupted (e.~g.~Sagittarius;   
Ibata, Gilmore \& Irwin 1994, 1995).

Furthermore, it can be possible to identifty groups of stars 
but be uncertain 
as to whether or not these stars form part
of a discrete galaxy with a dark halo.
Including or excluding such objects from the Local Group
galaxy sample can then be subjective.
An example is Andromeda VIII (Morrison et al.~2003), which may
be part of the M31 stream
(Ibata et al.~2004).
Another example is Canis Major (Martin
et al.~2004, Momany et al.~2004). 

        In Table 3 we present a list of Local Group members and their
        absolute $R$-band magnitudes.  The methods used to derive these
        magnitudes are: [1] interpolation using the spectral energy
        distributions (SEDs) of Fukugita et al.~(1995; hereafter F95), adopting
        a literature value for the absolute magnitude in some filter, and choosing a
        SED appropriate for the Hubble type of the galaxy; [2] as [1] but 
        choosing a SED appropriate for a galaxy with broadband colours
        equal to those measured for the galaxy; [3] as [1] but 
        choosing a SED appropriate for a galaxy with broadband colours
        equal to those inferred from colour-magnitude diagrams; [4]
        adopting a literature value for $M_R$.

        The LF for the Local Group is
        presented in Figure 4. The faint-end slope is $\alpha = -1.1
        \pm 0.1$, consistent with the findings of 
        Pritchet \& van den Bergh (1999) and van den Bergh (2000).

\begin{table} \caption{Local Group members}
{$$\vbox{
\halign {#\hfil && \quad #\hfil \cr
\noalign{\hrule \medskip}
Galaxy & type & $M_R$ & method & reference &\cr
\noalign{\smallskip \hrule \smallskip}
\cr
M31 = NGC 224 & Sb  & $-21.8$ & 1 & 1 &\cr
Milky Way     & Sbc & $-21.5$ & 1 & 2 &\cr
M33 = NGC 598 & Sc  & $-19.5$ & 3 & 2,3 &\cr
LMC & dI & $-18.8$ & 3 & 2,4 &\cr 
SMC & dI & $-17.4$ & 3 & 2,5 &\cr 
M32 = NGC 221 & E & $-17.1$ & 2 & 6,7 &\cr
NGC 205          & dE & $-16.9$ & 2 & 6,7 &\cr 
NGC 6822         & dI & $-16.5$ & 2 & 6,7 &\cr 
IC 10 & dI & $-16.3$ & 2 & 2,8 &\cr
NGC 3109         & dI & $-16.1$ & 2 & 6,7 &\cr 
NGC 185          & dE & $-15.7$ & 2 & 6,7 &\cr 
IC 1613          & dI & $-15.8$ & 3 & 6,9 &\cr
NGC 147          & dE & $-15.7$ & 2 & 6,7 &\cr 
Sagittarius      & dE & $-15.4$ & 3 & 6,10 &\cr
Sextans A        & dI & $-14.9$ & 3 & 6,11 &\cr
WLM              & dI & $-14.7$ & 3 & 6,12 &\cr
Sextans B        & dI & $-14.7$ & 3 & 6,13 &\cr
Fornax           & dE & $-13.6$ & 3 & 6,14 &\cr
Pegasus = DDO 216 & dI & $-13.4$ & 3 & 6,15 &\cr 
And VII = Cassiopeia& dE & $-12.7$ & 3 & 6,16 &\cr
Leo I            & dE & $-12.6$ & 3 & 6,17 &\cr
And I            & dE & $-12.5$ & 3 & 6,18 &\cr
And II           & dE & $-12.5$ & 3 & 6,18 &\cr
SagDIG           & dE & $-12.3$ & 4  & 19 &\cr
Leo A            & dI & $-12.1$ & 3 & 6,20 &\cr
Antlia           & dE & $-11.9$ & 3 & 6,21 &\cr
Aquarius = DD0 210 & dE & $-11.5$ & 3 & 6,22 &\cr
And VI = Pegasus & dE & $-11.1$ & 3 & 6,23  &\cr
LGS 3            & dI & $-10.9$ & 3 & 24,25 &\cr
And III          & dE & $-10.8$ & 3 & 18,26 &\cr
Cetus           & dE & $-10.7$ & 3 & 6,27 &\cr
Leo II           & dE & $-10.6$ & 3 & 6,28 &\cr
Sculptor         & dE & $-10.3$ & 3 & 6,29 &\cr
Phoenix         & dE/I & $-10.3$ & 3 & 6,30 &\cr
Tucana           & dE & $-10.2$ & 3 & 6,31 &\cr
Sextans          & dE & $-10.1$ & 3 & 26,32 &\cr
Draco            & dE & $-10.0$ & 3 & 6,33  &\cr
And V            & dE & $ -9.8$ & 3 & 34,35 &\cr
Carina           & dE & $ -9.9$ & 3 & 1,11,15 &\cr
Ursa Minor       & dE & $ -9.8$ & 3 & 34,35 &\cr
And IX           & dE & $ -8.8$ & 3 & 1,10  &\cr \noalign{\smallskip \hrule}
\noalign{\smallskip}\cr}}$$}
\begin{list}{}{}
\item[1.~]SED from NASA/IPAC Extragalactic Database 
(nedwww.ipac.caltech.edu); 
2.~van den Bergh 2000; 
3.~McConnachie et al.~2004;
4.~Castro et al.~2001;
5.~Alciano et al.~2003;
6.~Grebel et al.~2003;
7.~Prugniel \& Heraudeau 1998; 
8.~de Vaucouleurs \& Ables 1965; 
9.~Tikhonov \& Galazutdinova 2002;
10.~Ibata et al.~1997; 
11.~Dolphin et al.~2003;
12.~Rejkuba et al.~2000; 
13.~Sakai et al.~1997;
14.~Bersier \& Wood 2002;
15.~Gallagher et al.~1998;
16.~Grebel \& Guhatakurta 1999;
17.~Held et al.~2001;
18.~Da Costa, Armandroff \& Caldwell 2002;
19.~Lee \& Kim 2000;
20.~Dolphin et al.~2002;
21.~Tolstoy \& Irwin 2000;
22.~Lee et al.~1999;
23.~Hopp et al.~1999;
24.~Lee 1995a;
25.~Miller et al.~2001;
26.~Caldwell et al.~1992;
27.~Sarajadini et al.~2001;
28.~Lee 1995b;
29.~Monkiewicz et al.~1999;
30.~Gallart et al.~2001;
31.~Castellani, Marconi \& Buonanno 2002; 
32.~Bellazzini et al.~2001;
33.~Klessen, Grebel \& Harbeck 2003;
34.~Caldwell 1999;
35.~Davidge et al.~2002;
36.~Rizzi et al.~2003;
37.~Zucker et al.~2004 
\end{list}
\end{table}

\begin{figure}
\begin{center}
\vskip-4mm
\psfig{file=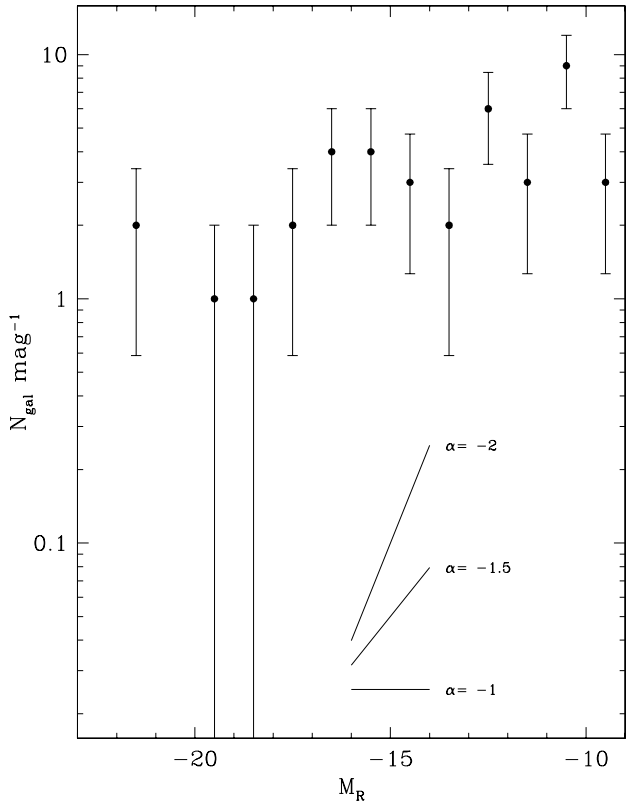, width=8.65cm}
\end{center}
\vskip-3mm
\caption{
The luminosity function of the Local Group.
}
\end{figure}

        Recently Karachentsev et al.~(2004) compiled a list of
        galaxies in the local Universe with distances less than 10
        Mpc, the Local Volume (LV) catalogue.
        This sample is probably incomplete at the very faint end at
        present but over the next few years it should be possible
	to identify and measure magnitudes and distances (by, for
	example, the tip of the red giant branch method) for the
   	vast majority of galaxies within 10 Mpc.  This will
	allow us to generate a LF with much smaller errors than in
	Figure 4.
 
\section{Combining the measurements -- the galaxy LF}

        The field luminosity function, obtained by combining the LFs
 	described in the previous three sections, is presented in
	Figure 5 and Table 4.  Prior to being combined, the LFs were
	normalized to a consistent scale and were weighted at each
	magnitude by the inverse of the square of the uncertainty. 	

\begin{figure}
\begin{center}
\vskip-4mm
\psfig{file=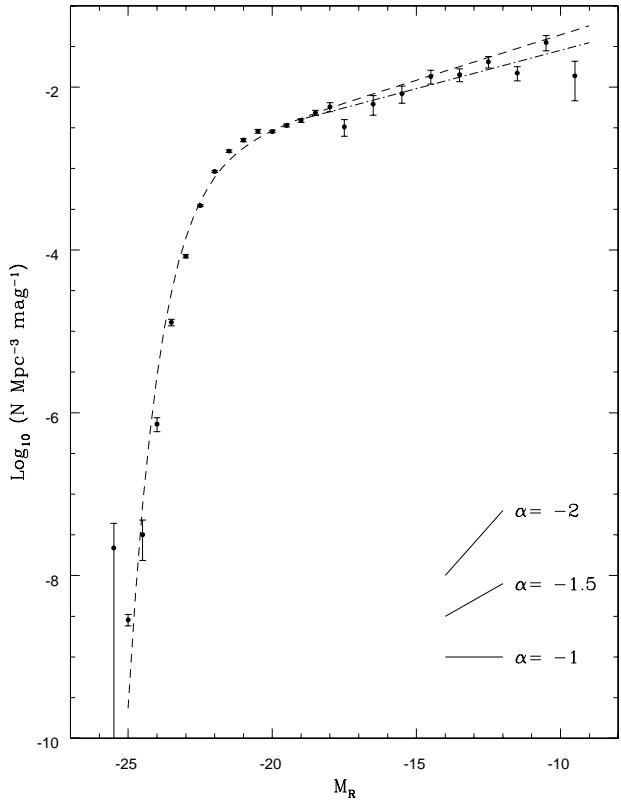, width=8.65cm}
\end{center}
\vskip-3mm
\caption{
The field galaxy luminosity function.
The normalization is as appropriate to the SDSS data and the other
samples are grafted onto the SDSS LF.  The dashed line
shows the best Schechter function fit to all the data: $M_R^* = -22.0$,
$\alpha^*=-1.29$.  The dotted-dashed line shows the best power-law fit for
$M_R > -19$: $\alpha=-1.26$.}
\end{figure}

\begin{table} \caption{The Field Galaxy Luminosity Function}
{$$\vbox{
\halign {#\hfil && \quad #\hfil \cr
\noalign{\hrule \medskip}
$M_R$ & $\phi$ &\cr
      & N mag$^{-1}$ Mpc$^{-3}$ &\cr
\noalign{\smallskip \hrule \smallskip}
\cr
$-25.5$ & $(2.18 \pm 2.18) \times 10^{-8}$ &\cr
$-25.0$ & $(2.86 \pm 0.46) \times 10^{-9}$ &\cr
$-24.5$ & $(3.17 \pm 1.64) \times 10^{-8}$ &\cr
$-24.0$ & $(7.26 \pm 1.40) \times 10^{-7}$ &\cr
$-23.5$ & $(1.30 \pm 0.12) \times 10^{-5}$ &\cr
$-23.0$ & $(8.38 \pm 0.36) \times 10^{-5}$ &\cr
$-22.5$ & $(3.50 \pm 0.091) \times 10^{-4}$ &\cr
$-22.0$ & $0.000920 \pm 0.000026$ &\cr
$-21.5$ & $0.00164 \pm 0.000052$ &\cr
$-21.0$ & $0.00224 \pm 0.000092$ &\cr
$-20.5$ & $0.00288 \pm 0.00015$ &\cr
$-20.0$ & $0.00285 \pm 0.000096$ &\cr
$-19.5$ & $0.00339 \pm 0.00014$ &\cr
$-19.0$ & $0.00390 \pm 0.00019$ &\cr
$-18.5$ & $0.00483 \pm 0.00035$ &\cr
$-18.0$ & $0.00573 \pm 0.00072$ &\cr
$-17.5$ & $0.00325 \pm 0.00075$ &\cr
$-16.5$ & $0.00694 \pm 0.0019$ &\cr
$-15.5$ & $0.00931 \pm 0.0022$ &\cr
$-14.5$ & $0.0150 \pm 0.0029$ &\cr
$-13.5$ & $0.0156 \pm 0.0028$ &\cr
$-12.5$ & $0.0228 \pm 0.0037$ &\cr
$-11.5$ & $0.0166 \pm 0.0033$ &\cr
$-10.5$ & $0.0393 \pm 0.0083$ &\cr
$-9.5$  & $0.0151 \pm 0.0073$ &\cr
\noalign{\smallskip}\cr}}$$}
\end{table}

        Figure 6 shows how much each individual LF contributes to the
	total LF.  The SDSS dominates the LF at the bright end and the
	CCD mosaic surveys at the faint end.  The Local Group is important 
	only for the faintest ($M_R = -9.5$) point.

\begin{figure}
\begin{center}
\vskip-4mm
\psfig{file=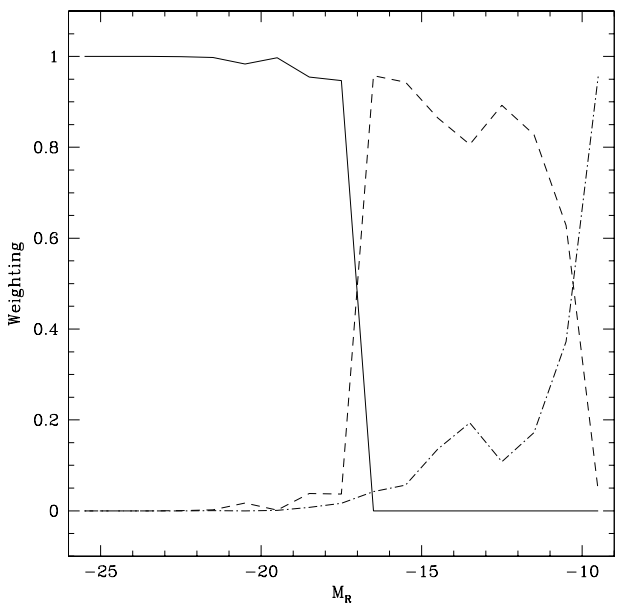, width=8.65cm}
\end{center}
\vskip-3mm
\caption{
The weighting of each component to the total LF for the SDSS (solid line),
CCD mosaic (dashed line) and Local Group (dotted-dashed line) samples.
The weightings here are proportional to $\sigma^2$, where $\sigma$ is the error
in the component LF.}
\end{figure}

Two points on the LF are worthy of further discussion: 
\vskip 1pt \noindent
(1) the $M_R=-17.5$ point.
This is the faintest point where the SDSS dominates and it is significantly
lower than all other points in this region.  One possibility is that the
SDSS sample is incomplete here beacuse galaxies with this
absolute magnitude have surface brightnesses too low for spectroscopy.
On the other hand,
the majority of galaxies with $M_R=-17.5$ ($M_B \sim -16.5$) have
surface brightnesses within their half-light-radius 
of $\mu_{50}\sim 20$ $R$ mag arcsec$^{-2}$ 
(Trentham, Tully \& Verheijen 2001b), which is bright enough for
spectroscopy with the instruments used by SDSS.  
Another possibility is that galaxies are
rare at precisely this magnitude, which 
correponds roughly to the
transition magnitude between giant and dwarf galaxies.
Interestngly, Flint, Bolte \& Mendes de Olivera (2003) report a
deficiency in galaxies with absolute magnitudes close to this value in the
Leo I Group.
\vskip 1pt \noindent
(2) the $M_R=-9.5$ point.
There are only three Local
Group galaxies which contribute to this point, and all
narrowly escaped inclusion into the $M_R=-10.5$ bin.  
While completeness and small number statistics
are surely concerns at this extreme faint end,
this observation may be a hint of a turnover in the LF at about
$M_R=-9.5$.  Galaxies like Andromeda IX would then be extremely rare.
The shape of the LF here and the possible existence of a turnover
should become well-established over the next few years as the LV sample
is extended to larger distances 
and completeness issues in that sample become better understood.

        The normalization of the LF is affected
	by cosmic variance and will vary from survey to
	survey.   The value adopted in Figure 5 comes from SDSS
	observations which cover 0.3 \% of the sky and 
	so should provide a reasonable representation of the cosmological
	average.

\section{analytic forms}

A power-law fit to the LF over the range $M_R > -19$ gives a logarithmic
slope $\alpha = -1.2$ but this is not a good fit in that it
overpredicts the number of galaxies at the extreme faint end.  The
error bars in Figure 5 are small enough that curvature in the LF
within this magnitude range is detectable at a high level of
significance.

It is common to fit LFs with Schechter (1976) functions:
$$\phi (L) = \phi^* \exp(-{L\over{L^*}}) 
\left({L\over{L^*}}\right)^{\alpha}\,{1\over{L^*}},\eqno(1)$$
or in magnitude units
$$\phi(M) = 
% - {{{\rm d} L }\over{{\rm d} M }} \phi(L) =$$ 
0.92 \,
\phi^* \, 
(10^{[-0.4(M-M^*)]})^{\alpha^* + 1} \, 
{e}^{-{10^{[-0.4(M-M^*)]}}}.\eqno(2)$$
Here $\phi^*$ is a normalization density,
$L^*$ and $M^* = -2.5 \, \log_{10} L^*$ are 
a characteristic luminosity and magnitude, and
$\alpha^*$ is a characteristic faint-end slope.
These fits normally have $\alpha > -2$
and so provide convenient ways to describe galaxy
luminosities over a magnitude range
(close to $M^*$) where most of the total luminosity of a sample of
galaxies is obtained.
Schechter function fits have no physical basis
but are attractive in that their 
exponential cutoff is similar to that predicted by simple analytical
models of galaxy formation (e.~g.~White \& Rees 1978).  

A Schechter function fit over the entire range $-25 < M_R
< -9$ is poor because (1) it is not possible to simultaneously
fit the data at both $M_R = -22$ and $M_R = -20$ where
the errors in the SDSS LF are so small,
(2) the point at $M_R = -17.5$ is systematically too low, and
(3) a Schechter function which fits the data well at bright 
magnitudes then overpredicts the number of faint galaxies.  
These features are illustrated by the dashed line on Figure 5.

Two-parameter fits cannot fit the data over any significant 
magnitude range.
Composite forms, like linear combinations of Gaussian and
Schechter functions, may provide acceptable fits to
the data but there is
no physical basis for selecting a particular set of
functions at this stage.  

\section{Comparison with theory}

It is well known that cold dark matter (CDM) theory 
predicts a mass function of dark halos 
that is steep at the low-mass 
end (e.~g.~Moore 
et al.~1999).  
Simulations predict a logarithmic mass function slope
$\alpha \sim -2$ at low masses.

The field galaxy LF is much shallower than this at all 
magnitudes.  This tells us that star formation must be less
efficient in lower-mass galaxies.  There are various physical
processes that can be responsible for this.
Lower-mass galaxies are more susceptible to
gas ejection by supernova-driven winds (Dekel \& Silk 1986)
due to their having smaller gravitational potential wells.
This means that negative feedback effects during star formation
(Efstathiou 2000) are stronger in lower-mass galaxies.
Additionally, in hierarchical clustering cosmologies like CDM, 
many low-mass
galaxies form at later times so it will be difficult
for them to collect gas from an intergalactic medium that is being
progressively heated by the metagalactic ultraviolet background
(Thoul \& Weinberg 1996;
Klypin et al.~1999; Bullock,
Kravstov \& Weinberg 2001;
Tully et al.~2002).  

Whatever processes are responsible, we shall
see below that they must operate less efficiently in galaxy clusters. 
Perhaps the presence of a confining medium (Babul \& Rees 1992)
inhibits the effects of negative feedback during low-mass galaxy
formation (see Section 5 of Roberts et al.~2004).

\section{the cluster LF}

Various measurements of cluster LFs have been made possible by a
number of advances in technology.

The prototypical rich galaxy cluster is the Coma cluster, with
several hundred galaxies brighter than $M_R = -19$ per
square degree.  
Recently a spectroscopic LF of Coma extending down to $M_R = -16$
has been published (Mobasher et al.~2003).  This is a difficult
measurement to make because it requires spectroscopy of large
numbers of faint sources, many of which have relatively low
surface brightnesses.  This measurement has been made possible by the advent of
wide-field multi-object spectrographs that can be used on large
telescopes, in this case WYFFOS on the 4.2 m William Herschel
Telescope (Bridges 1998).
  
Many other clusters that are richer and more distant than Coma
have also been studied (A665 at $z=0.18$: 
the only Abell (1958) Richness 5 cluster,
Wilson et al.~1997, Trentham 1998; 
A963 at $z=0.21$, Driver et al.~1994, Trentham 1998;
A1689 at $z=0.18$, Wilson et al.~1997; 
A868 at $z=0.15$, Boyce et al.~2001; 
A1146 at $z=0.14$, Trentham 1997).
These clusters have very well-determined
LFs at the bright end because their contrast against the background field
galaxy population is so
high.  Photometry alone can therefore be used for a 
measurement of the LF at bright absolute
magnitudes.  When many such clusters are considered
in conjunction with each other, the numbers of galaxies in each bin
becomes large and the Poisson errors are consequently small.
At the faint end, these measurements cannot be used to
determine the LF since the cluster counts are small compared to
the background counts and the field-to-field variance of the
background is much larger than the cluster signal.  Spectroscopic
measurements cannot help either at the faint end since the clusters are
so distant. 

The Virgo cluster is much less rich than the Coma Cluster, but 
at 17 Mpc it is
sufficiently nearby that dwarf galaxies are
much larger than the seeing and can be identified by
their low surface brightnesses.
The LF can then be determined to
faint limits by photometric observations alone. 
This LF will not include the contribution from compact galaxies like M32
and the Ultra Compact Dwarfs in the Fornax Cluster (Phillipps 
et al.~2001, Drinkwater et al.~2003)
because they will be mis-identified as background ellipticals.
However these types of galaxies are rare.
The Virgo Cluster Catalog (Binggeli, Sandage \& Tammann 1985) 
was compiled about 20 years ago and
provided a precise measurement of the bright end of Virgo Cluster
LF (Sandage, Binggeli \& Tammann 1985).  More recently, mosaic
CCD measurements (Trentham \& Hodgkin 2002, Sabatini et al.~2003)
using the Wide Field Camera on the 2.5 m Isaac Newton
Telescope (INT) have confirmed the earlier results and have extended
the LF faintward by about 5 magnitudes.  Extremely deep observations
over a smaller area of Virgo using
SuprimeCam on the 8 m Subaru Telescope (Trentham \& Tully 2002)
did not result in the discovery of any new
dwarfs brighter than the completeness limit of the INT survey
(about $M_B=-11$) that were too low in surface brightnesses to be seen
with the smaller telescope. 
The implication is then 
that the LF of the Virgo Cluster is known at least as faint as 
$M_B=-11$.

We derive a composite cluster LF by combining measurements from
photometric studies of
rich galaxy clusters (A665, A963, A1795, A2199 and A1146; Trentham 1997, 1998),
the deep spectroscopic study of the Coma cluster (Mobasher et al.~2003), and a
photometric study of the Virgo cluster (Trentham \& Hodgkin 2002).  
Each dataset is important in a different absolute magnitude regime 
(Figure 7).  At very bright absolute magnitudes, the rich cluster sample
is most important; the Coma and Virgo samples suffer from small number
statistics.  At intermediate magnitudes, the Coma sample is most
important; the rich cluster LF has large errors since a background subtraction
is required and the background has a large field-to-field variance and
the Virgo sample has substantial Poisson errors.
At faint magnitudes only the Virgo data can be used; the other surveys
cannot detect dwarfs, most of which have low surface brightnesses. 
Although the three samples utilise very different methods in establishing
membership,
they are 
consistent with each other in the absolute magnitude 
where they overlap. 
All data used are $R$-band data except the
Virgo data which are $B$-band data converted to the $R$-band
using a combination of (1) $B-R$ colours from a calibration
derived using measurements 
of faint galaxies 
observed in both the INT 
and Subaru surveys,
(2) the spectral energy distributions from F95, and (3) the $B-R$ calibration
for giant galaxies of different Hubble Type $T$
obtained using data from Tully \& Pierce (2000):
$B-R=1.40 - 0.059 T$ (equation (1) in TT02).
Typical random errors in this conversion are estimated to be 0.1 -- 0.2 mag, 
smaller than the bin size of 0.5 mag.

\begin{figure}
\begin{center}
\vskip-4mm
\psfig{file=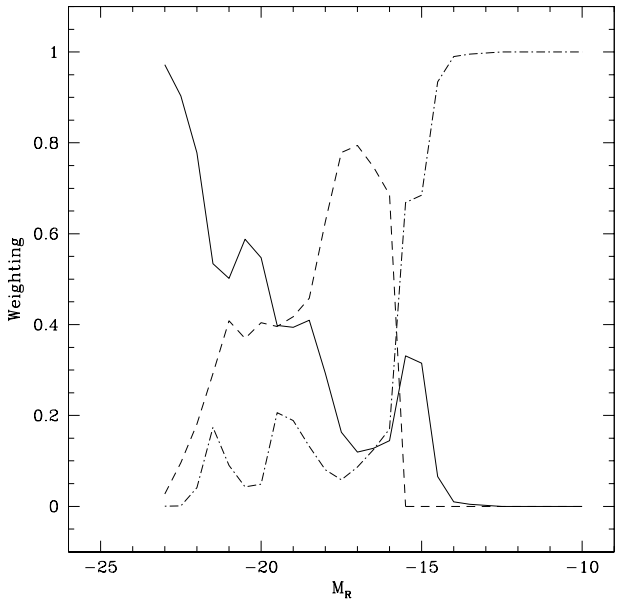, width=8.65cm}
\end{center}
\vskip-3mm
\caption{
The weighting of each component to the total LF.
The lines are for the rich cluster
(solid line),
Coma (dashed line) and Virgo (dotted-dashed line) samples.
The weightings here are proportional to $\sigma^{-2}$, 
where $\sigma$ is the error
in the component LF.}
\end{figure}

The cluster LF is presented in Table 5 and Figure 8. 
This LF is appropriate for centres of galaxy clusters although the
inclusion or exclusion of cD galaxies in the samples used to define the
LF is arbitrary.  There is evidence, however, that the LF might be quite
different in the outer parts of galaxy clusters (Phillipps et al.~1998).

\begin{figure}
\begin{center}
\vskip-4mm
\psfig{file=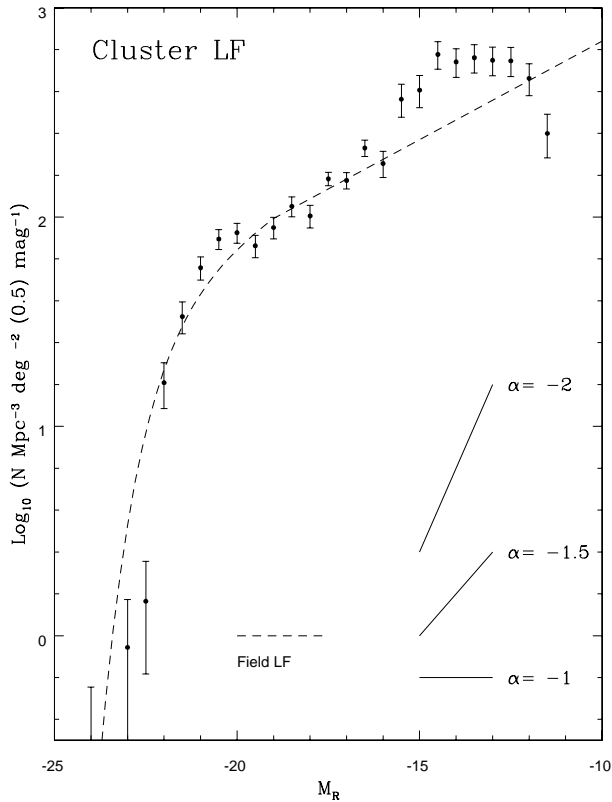, width=8.65cm}
\end{center}
\vskip-3mm
\caption{
The galaxy cluster luminosity function, computed as described as in the
text.  The normalization is arbritrarily chosen to be that appropriate for
the Coma Cluster at $M_R=-21$.  The dashed line is the best-fitting
field luminosity function, which we
approximate by a Schechter function with $M_R^* = -22.0$ and
$\alpha^*=-1.28$ brightward of $M_R=-19$ and a power law with
$\alpha=-1.24$. faintward of $M_R=-19$.}
\end{figure}

\begin{table} \caption{The Galaxy Cluster Luminosity Function}
{$$\vbox{
\halign {\hfil #\hfil && \quad \hfil #\hfil \cr
\noalign{\hrule \medskip}
$M_R$ & $\phi$ &\cr
      & N mag$^{-1}$ Mpc$^{-3}$ &\cr
\noalign{\smallskip \hrule \smallskip}
\cr
$-23.0$ & $0.88 \pm 0.61$ &\cr
$-22.5$ & $1.46 \pm 0.80$ &\cr
$-22.0$ & $16.2 \pm 4.0$ &\cr
$-21.5$ & $33.4 \pm 5.8$ &\cr
$-21.0$ & $57.3 \pm 7.2$ &\cr
$-20.5$ & $78.5 \pm 8.6$ &\cr
$-20.0$ & $84.2 \pm 9.2$ &\cr
$-19.5$ & $72.9 \pm 8.9$ &\cr
$-19.0$ & $89.1 \pm 10.4$ &\cr
$-18.5$ & $113 \pm 12$ &\cr
$-18.0$ & $101 \pm 13$ &\cr
$-17.5$ & $152 \pm 11$ &\cr
$-17.0$ & $150 \pm 14$ &\cr
$-16.5$ & $214 \pm 19$ &\cr
$-16.0$ & $180 \pm 26$ &\cr
$-15.5$ & $366 \pm 65$ &\cr
$-15.0$ & $404 \pm 71$ &\cr
$-14.5$ & $599 \pm 89$ &\cr
$-14.0$ & $551 \pm 87$ &\cr
$-13.5$ & $577 \pm 89$ &\cr
$-13.0$ & $561 \pm 88$ &\cr
$-12.5$ & $558 \pm 88$ &\cr
$-12.0$ & $460 \pm 80$ &\cr
$-11.5$ & $250 \pm 59$ &\cr
\noalign{\smallskip}\cr}}$$}
\end{table}

The cluster LF can be fit by
a Schechter function brightward of $M_R=-16$, where
it rises with a logarithmic slope $\alpha=-1.6$ and then is flat
faintward of about $M_R=-13$. 
The faintest point on the LF shown in Figure 8 might be affected by
incompleteness (although the deep Subaru 
Virgo measurements suggest
otherwise) but the flattening faintward of about $M_R=-13$
comes from an analysis of the Virgo data in an absolute magnitude
regime where the completeness is close to 100 per cent.
 
The field and cluster LFs are highly inconsistent with each other
(Figure 9).
Brighter than $M_R=-20$, the LF falls off slightly more steeply in the field
than in clusters due to the presence of luminous elliptical galaxies in clusters.
At about $M_R=-20$, the field and cluster LFs have similar shapes,
despite being dominated by contributions from different kinds of
galaxies (early-type galaxies in clusters and late-type galaxies
in the field).
The cluster LF becomes significantly steeper than the field LF at about
$M_R=-17$ due to the
presence of a large number of dwarf elliptical galaxies.  Both LFs are
reasonably flat faintward of about $M_R=-15$, 
although the cluster LF has a higher normalization relative to the giant
galaxies than does the field LF and the contribution of dwarf irregular
galaxies relative to that of dwarf elliptical galaxies is higher in
the field than in clusters.
This excess number of cluster relative to field dwarfs is described and
studied in detail by Roberts et al.~(2004), although the differential
between the cluster and field that we find is less than that found by
those authors.
This may be due in part due to there being slightly fewer dwarfs per
giant in groups than in environments with very low galaxy densities
(see Zabludoff \& Mulchaey 2000).

\begin{figure}
\begin{center}
\vskip-4mm
\psfig{file=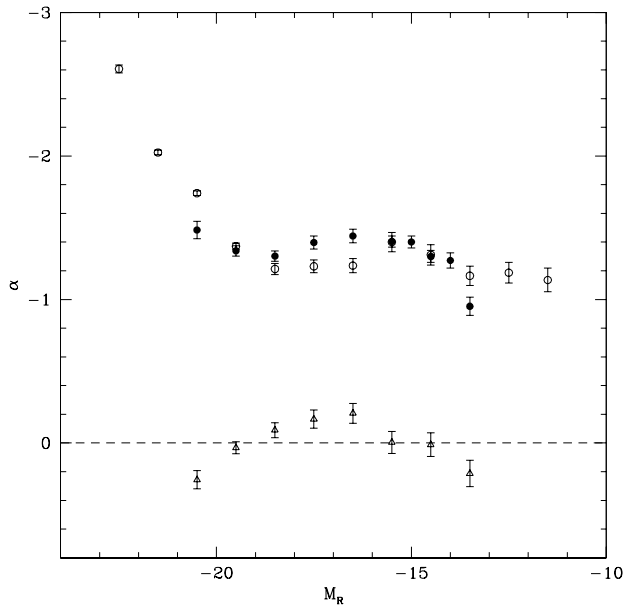, width=8.65cm}
\end{center}
\vskip-3mm
\caption{
The logarithmic slope $\alpha$ of
the field (open circles) and 
cluster (filled circles) LFs, and the difference between the two
(open triangles).  The points are calculated at each
absolute magnitude $M_R$ by considering the shape of the LF over a
two magnitude interval centred on $M_R$.}
\end{figure}

The faint end of the cluster luminosity function depends on
measurements of the Virgo Cluster and it is important to
establish that this environment is representative -- different
studies of the faint end of the Virgo LF (Trentham \& Hodgkin 2002,
Trentham \& Tully 2002, Sabatini et al.~2003) concur as regards the
general shape (rise at $M_R=-17$ and flattening at fainter magnitudes)   
but comparisons betweeen different clusters at the very faint end
have yet to be made.
There is a hint of a LF with a similar shape in the knot of early-type
galaxies around NGC 1407 (Trentham \& Tully 2002) but this sample is
small.  The best place to determine a LF that can be compared with the
Virgo LF is the Fornax Cluster, which also has a population of
low-surface-brightness dwarf galaxies (Kambas et al.~2000).
Spectroscopic information is available for a subset (about one-quarter)
of the Fornax Cluster Spectroscopic Survey (FCSS) dwarfs,  
but there are too few dwarfs known (19 with $-15.8 < M_B < -12.7$; Deady
et al.~2002) for a LF to be computed that can be compared in detail
with the Virgo LF.   
When the FCSS is complete, this will be an important
comparison.

The previous paragraph is concerned with the precise shape
of the LF at the faintest magnitudes.
All current evidence does at this stage points towards an excess of 
low-luminosity galaxies in clusters, although this precise shape
depends on the measurements of the local clusters.
For example, an excess of low luminosity cluster 
galaxies above the expected number
given the field LF was seen in the 2dF redshift survey 
(de Propris et al.~2003).

This tendency for dwarfs to be more numerous in the richer
and denser environments
is important in that it tells us that the physical processes 
responsible for suppressing the formation of stars in low-mass dwarf galaxies
may operate less
efficiently in clusters than in the field.  Alternatively there
are additional physical processees that operate preferentially in clusters
and alleviate the effects of dwarf suppression there.  For a more detailed 
discussion of these issues, the reader is referred to Roberts et al.~(2004). 

The shape of the cluster LF is such that it cannot be
described by a simple analytic form.  Composite forms may be more
appropriate; these may provide a useful test of theories which may wish
to invoke different formation mechanisms for giant and dwarf galaxies.
For example, the cluster LF is well fit by a double Schechter function
with $(M^*,\alpha^*) = (-21.4, -1.0)$ for giants and
$(M^*,\alpha^*) = (-16.6, -1.1)$ for dwarfs and a relative normalization
of $2.7 \, \times 10^{-0.4 \, (M^*_{\rm giant} - M^*_{\rm dwarf})}$. 

\section{Summary}

This paper presents the
results of a study that combines different datasets in order to measure
the field galaxy LF over the range $-25 < M_R < -9$.
The data come from the
Sloan Digital Sky Survey, from CCD mosaic imaging surveys of nearby groups, and
from the Local Group.
The field luminosity function
is well-described by a Schechter function with $M_R^* = -22.0$ and
$\alpha^*=-1.28$ brightward of $M_R=-19$ and by a power law with
$\alpha=-1.24$ faintward of $M_R=-19$.
This is shallower than the CDM mass function ($\alpha \sim -2$), 
implying that star formation is less efficient in lower-mass
galaxies.

A similar analysis was performed to measure the LF of galaxy clusters.
The three samples used were a photometric survey of rich Abell clusters,
a spectroscopic survey of the Coma Cluster and a
photometric survey of the Virgo Cluster.
The cluster LF differs from the field LF in that it has a rise of $\alpha \sim
-1.6$ between $M_R = -17$ and $M_R = -14$, although it flattens
faintward of $M_R = -14$.
In the context of CDM theories, 
this suggests that the physical processes that are
responsible for suppressing the formation of stars in low-mass galaxies
operate less
efficiently in clusters than in the field. 

\section*{Acknowledgements}
This research has made use of the NASA/IPAC Extragalactic Database (NED)
which is operated by the Jet Propulsion Laboratory, Caltech, under agreement
with the National Aeronautics and Space Association.

\end{document}